\documentclass[12pt]{article}
\usepackage[margin=1in]{geometry}
\usepackage{setspace}

\usepackage{float}

\usepackage{amsfonts}

\usepackage{amssymb}
\usepackage{latexsym}
\usepackage{graphicx}
\usepackage[english]{babel}
\usepackage[font={small}]{caption}

\usepackage{amsfonts}
\usepackage{latexsym}
\usepackage{graphicx}
\usepackage[english]{babel}
\usepackage{tikz}

\begin{document}
\input epsf

\def\p{\partial}
\def\h{{1\over 2}}
\def\be{\begin{equation}}
\def\bea{\begin{eqnarray}}
\def\ee{\end{equation}}
\def\eea{\end{eqnarray}}
\def\d{\partial}
\def\la{\lambda}
\def\eps{\epsilon}
\def\bb{\bigskip}
\def\mm{\medskip}
\newcommand{\dm}{\begin{displaymath}}
\newcommand{\edm}{\end{displaymath}}
\renewcommand{\b}{\tilde{B}}
\newcommand{\gm}{\Gamma}
\newcommand{\ac}[2]{\ensuremath{\{ #1, #2 \}}}
\renewcommand{\ell}{l}
\newcommand{\z}{\ell}
\newcommand{\newsection}[1]{\section{#1} \setcounter{equation}{0}}
\def\bb{$\bullet$}
\def\Qbar{{\bar Q}_1}
\def\QPbar{{\bar Q}_p}

\def\q{\quad}

\def\bn{B_\circ}

\let\a=\alpha \let\b=\beta \let\g=\gamma \let\d=\delta \let\e=\epsilon
\let\c=\chi \let\th=\theta  \let\k=\kappa
\let\l=\lambda \let\m=\mu \let\n=\nu \let\x=\xi \let\r=\rho
\let\s=\sigma \let\t=\tau
\let\vp=\varphi \let\vep=\varepsilon
\let\w=\omega      \let\G=\Gamma \let\D=\Delta \let\Th=\Theta
                     \let\P=\Pi \let\S=\Sigma

\def\h{{1\over 2}}
\def\t{\tilde}
\def\r{\rightarrow}
\def\nn{\nonumber\\}
\let\bm=\bibitem
\def\Kt{{\tilde K}}
\def\b{\bigskip}
\def\m{\medskip}

\let\p=\partial

\newcommand\blfootnote[1]{%
  \begingroup
  \renewcommand\thefootnote{}\footnote{#1}%
  \addtocounter{footnote}{-1}%
  \endgroup
}

\newcounter{daggerfootnote}
\newcommand*{\daggerfootnote}[1]{%
    \setcounter{daggerfootnote}{\value{footnote}}%
    \renewcommand*{\thefootnote}{\fnsymbol{footnote}}%
    \footnote[2]{#1}%
    \setcounter{footnote}{\value{daggerfootnote}}%
    \renewcommand*{\thefootnote}{\arabic{footnote}}%
    }

\begin{flushright}
\end{flushright}
\vspace{6mm}
\begin{center}
{\LARGE  Approaching the surface of an Exotic Compact Object\daggerfootnote{Essay awarded 4th prize in the Gravity Research Foundation 2026 Awards for Essays on Gravitation.}
 }

\vspace{18mm}
{\bf Shokoufe Faraji$^{a,c}$ and Samir D. Mathur$^{b,c}$ }

\vspace{8mm}

${}^a$ Department of Physics and Astronomy, University of Waterloo, 

Waterloo,
ON N2L 3G1, Canada

email: s3faraji@uwaterloo.ca

\b

${}^b$ Department of Physics, The Ohio State University
 Columbus,
OH 43210, USA

email: mathur.16@osu.edu

\b

${}^c$ Perimeter Institute for Theoretical Physics, Waterloo, 
ON N2L 2Y5, Canada

\b

\b


\vspace{4mm}
\end{center}
\vspace{8mm}
\thispagestyle{empty}
\begin{abstract}

Many approaches to quantum gravity require replacing the traditional black hole geometry with an Exotic Compact Object (ECO), which has a large but not infinite redshift at its surface. We argue that near the ECO surface,  the vacuum Einstein equations imply a metric that is chaotic, with increasingly large oscillations as we approach the surface. This behavior is analogous to the `cosmic billiards' found in the BKL analysis of cosmology near the big bang.  For the ECO, some of the potential walls of this  billiards change sign  to become `cliffs',  resulting in a runaway behavior where some compact directions squeeze to zero size.  In string theory such squeezing yields a  natural continuation to the interior geometry of fuzzballs, where compact directions collapse to create monopoles.

\end{abstract}
\vskip 1.0 true in

\newpage
\setcounter{page}{1}

\doublespace

Imagine that you are watching a science fiction film, where the villain falls into a very large black hole. His spaceship feels nothing special as it crosses the horizon, and heads to the central singularity. Near this singularity, longitudinal directions stretch enormously while transverse directions compress by a similar amount.  The ship is crushed, to cheers from the audience.

But this classical picture of the hole is in conflict with quantum theory,  as shown by Hawking's black hole information paradox \cite{hawking}. Hawking's argument has now been solidified to a precise contradiction  using the strong subadditivity of quantum entanglement entropy \cite{cern}. In string theory this contradiction is avoided because  black hole microstates  turn out to be `fuzzballs': horizon sized quantum stars  that have no horizon and radiate from their surface like normal bodies \cite{fuzzballs}.  Many other approaches have also  suggested that the traditional black hole be replaced by an Exotic Compact Object (ECO) \cite{eco}. In this case, what would be the fate of the spaceship headed towards the hole?

It is tempting to think that falling towards an ECO would be just like falling towards the moon: feeling nothing but the weak tidal force of gravity till we reach the surface, and then a hard impact whose details depend on the composition of the  surface.  But as we will argue in this essay, the situation is far more interesting for infall towards an ECO that is very compact; i.e.,  the redshift on its surface is $z_s\gg1$. As an infaller approaches the surface of the ECO, he gets stretched sharply along one axis while getting compressed along another.  A little while later his experience suddenly changes:  there are again axes of stretching and compression, but these axes are completely different from the earlier ones. These changes of axes occur more and more rapidly and sharply as he gets closer to the ECO surface.  The infaller thus gets kneaded like pizza dough, until he enters a final phase where the strength of the distortion heads towards infinity. This phase of distortion ends when he reaches the quantum gravitational physics of the ECO interior. What is remarkable is that this entire kneading process is  a consequence of just the vacuum Einstein equations   $R_{ab}=0$, assumed to hold outside the ECO surface.

\b

{\bf The BKL hypothesis}

\b

To understand the essence of this story, we must go back to an important debate in the 1960s and 70s, about the nature of singularities in general relativity. It had already been observed  that curvature singularities appeared in spherically symmetric gravitational collapse, and in the homogeneous and isotropic Robertson-Walker cosmologies.  But could such singularities be an artifact of the high symmetry assumed in these solutions? One view, championed by the Soviet school in particular, held that more generic initial conditions would lead to geodesics `missing' the singularity and emerging unscathed to another smooth region of spacetime. This possibility was eventually ruled out, however, by the singularity theorems of Penrose and Hawking, who showed that geodesics inside a trapped surface can continue only for a finite duration of their affine parameter. Thus we cannot evade a singularity either at the black hole center or at the big bang \cite{hawkingellis}.

But the singularity theorems say nothing about the nature of the singularity. The Soviet school -- led by Belinski, Khalatnikov and Lifshitz (BKL) -- then suggested that the singularity would in general have a very complex structure \cite{bkl}. Consider the case of cosmology, and assume that there is no matter. If we assume homogeneity and isotropy, then the only solution is Minkowski space $ds^2=-dt^2+dx_1^2+dx_2^2+dx_3^2$. But if we relax the requirement of isotropy, we find the Kasner solutions 
\be
ds^2=-dt^2+ t^{2p_1} dx_1^2 + t^{2p_2} dx_2^2+t^{2p_3} dx_3^2\, .
\label{one}
\ee
The Einstein equations impose two constraints on the Kasner indices $p_a$
\be
p_1+p_2+p_3=1, ~~~~p_1^2+p_2^2+p_3^2=1
\label{two}
\ee
which imply that one of the indices satisfies $p\le 0$ and two satisfy $p\ge 0$.  A direction with $p<0$ expands to infinite size as we approach the big bang, while directions with $p>0$ contract to zero size. 

But a generic solution will not satisfy the requirement of homogeneity either. In a seminal paper \cite{bkl},  BKL postulated that the effect of small inhomogeneities could be captured by making a simple approximation. Near the big bang, time derivatives of the metric components are typically very large, much larger than the space derivatives. Thus in general  the neighborhood of  each spatial point $x$ on a spacelike slice decouples from other regions on the slice. The metric around $x$ evolves approximately in the manner (\ref{one}), but with the Kasner indices $p_a$ and the Kasner axes $\{ x_1,x_2,x_3\} $ varying from region to region along the slice.  

But there is an important additional twist to this story. As we follow the metric at a spatial point $x$ back towards $t=0$, once in a while the space derivatives of the metric {\it do} become important. Their effect is to sharply change the Kasner indices $p_a$ and the Kasner axes $\{ x_1, x_2, x_3\} $ at the point $x$. The backwards evolution in $t$  then continues in the Kasner manner (\ref{one}), till we meet another sharp change in the Kasner variables.  This bouncing around of the indices can be recast as a `cosmic billiards', where the evolution of the metric coefficients follows an ergodic billiard ball motion between a set of confining walls. Thus the approach to the singularity is chaotic, both in space and in time. 

To see this in more detail, consider the following ansatz for a small neighborhood of the point $x=(0,0,0)$
\be
ds^2=-dt^2 +e^{2\beta_1(t)}(dx_1+C^1_{23} x_2 dx_3)^2+ e^{2\beta_2(t)} dx_2^2+e^{2\beta_3(t)} dx_3^2
\label{three}
\ee
where $C^1_{23}$ is a small constant. It is convenient to define a new time coordinate $u$ through
\be
dt= e^{\beta_1+\beta_2+\beta_3}\, du
\label{four}
\ee
 and to denote derivatives $\p_u$ by a prime. If we ignore $C^1_{23}$, then the vacuum Einstein equations give $\beta''_a= 0$, and we recover the Kasner solutions (\ref{one}). Keeping $C^1_{23}$ gives
 \be
 \beta''_1= \left (C^1_{23}\right)^2 e^{4\beta_1}, ~~~ \beta''_2= -\left (C^1_{23}\right)^2 e^{4\beta_1}, ~~~ \beta''_3= -\left (C^1_{23}\right)^2 e^{4\beta_1}\, .
 \label{five}
 \ee
 In the time intervals where the  RHS of these equations remains small, the $\beta_a$ drift with approximately constant $u$-velocities. But this drift can carry the system to a point where $e^{4\beta_1}$ is large, and then the RHS of these equations is {\it not} ignorable. The exponentially rising potential $Exp[4\beta_1]$ acts like a wall which reflects the velocities $\beta'_a$, leading to a new Kasner type epoch, with new Kasner indices
 \be
p'_1=-{p_1\over 1+2p_1}, ~~~~~p'_2={p_2+2p_1\over 1+2p_1}, ~~~~~p'_3={p_3+2p_1\over 1+2p_1}\, .
\label{six}
\ee
In a generic metric that is not exactly homogeneous and isotropic, there are small but nonvanishing  coefficients $C^a_{bc}$ for all choices of the spacelike indices $a,b,c$. The corresponding potential walls trap the coefficients $\beta_a$ into a billiard-type motion that is ergodic over the space of possible Kasner index sets $p_a$,  for all initial conditions except a set of measure zero. Interestingly,  these reflecting walls can be mapped to the Weyl reflections of a Kac-Moody algebra \cite{cosmic}.

\b

{\bf BKL-type dynamics near the ECO surface}

\b

To see how the above dynamics might say something about the structure of an ECO,  let us ask where Minkowski space  fits in the above discussion.  Using the coordinates of the 1+1 dimensional Milne universe:  $ T=t \cosh x_1, ~X=t \sinh x_1$,  we find
\be
-dT^2+dX^2+dx_2^2+dx_3^2=-dt^2+t^2dx_1^2+dx_2^2+dx_3^2
\label{ten}
 \ee
 so a patch of Minkowski space has the Kasner form (\ref{one}) with   $p_1=1, ~p_2=p_3=0$. 
 
Now consider an extremely compact ECO, with surface at  radius $R_{ECO}=2GM+\epsilon$ ($\epsilon\ll GM$). The near surface region is described by the Rindler metric
\be
ds^2= d\rho^2-\rho^2 dt^2 + dx_1^2+dx_2^2
\label{el}
\ee
which has  the form of the Kasner metric in (\ref{ten}), but with the role of space and time interchanged. More generally, we find that the vacuum Einstein equations $R_{ab}=0$ have the following time-independent analogues of the Kasner solutions (\ref{one})
\be
ds^2=d\rho^2+ \rho^{2p_1} dx_1^2 + \rho^{2p_2} dx_2^2-\rho^{2p_3} dt^2
\label{oneq}
\ee
with the same constraints (\ref{two}) on the indices $p_a$. 

For a simple example of such Kasner-type behavior, consider the  `q-metrics', which describe   non-spherically-symmetric vacuum geometries without  angular momentum \cite{qmetric}
\bea
ds^2&=&-(1-{2M\over r})^{1+q}dt^2 \nn
&+&{1\over (1-{2M\over r})^q}\left [ \left (1+{M^2\sin^2\theta\over r^2(1-{2M\over r})}\right )^{-q(2+q)}({dr^2\over 1-{2M\over r}}+r^2d\theta^2)+r^2\sin^2\theta d\phi^2\right ]
\label{twq}
\eea
Let us fix $\theta, \phi$, and look at the region $r'\equiv r-2M \ll M$.  Define a new radial variable $\rho$ through $r'\sim  \rho^{2\over q^2+q+1}$. Then the geometry takes the Kasner form (\ref{oneq}) with
\be
p_\theta={q^2+q\over q^2+q+1}, ~~~p_\phi=-{q\over q^2+q+1}, ~~~p_t={q+1\over q^2+q+1}\, .
\label{thir}
\ee
Setting $q=0$ recovers the Schwarzschild geometry  with the Kasner indices of Rindler space $p_\theta=p_\phi=0, \, p_t=1$.  (The metric (\ref{twq}) is singular at $r=2M$, but in our ECO some quantum gravitational behavior will take over before we reach this singular surface.) 

 \begin{figure}
  \center{   \includegraphics[scale=.16]{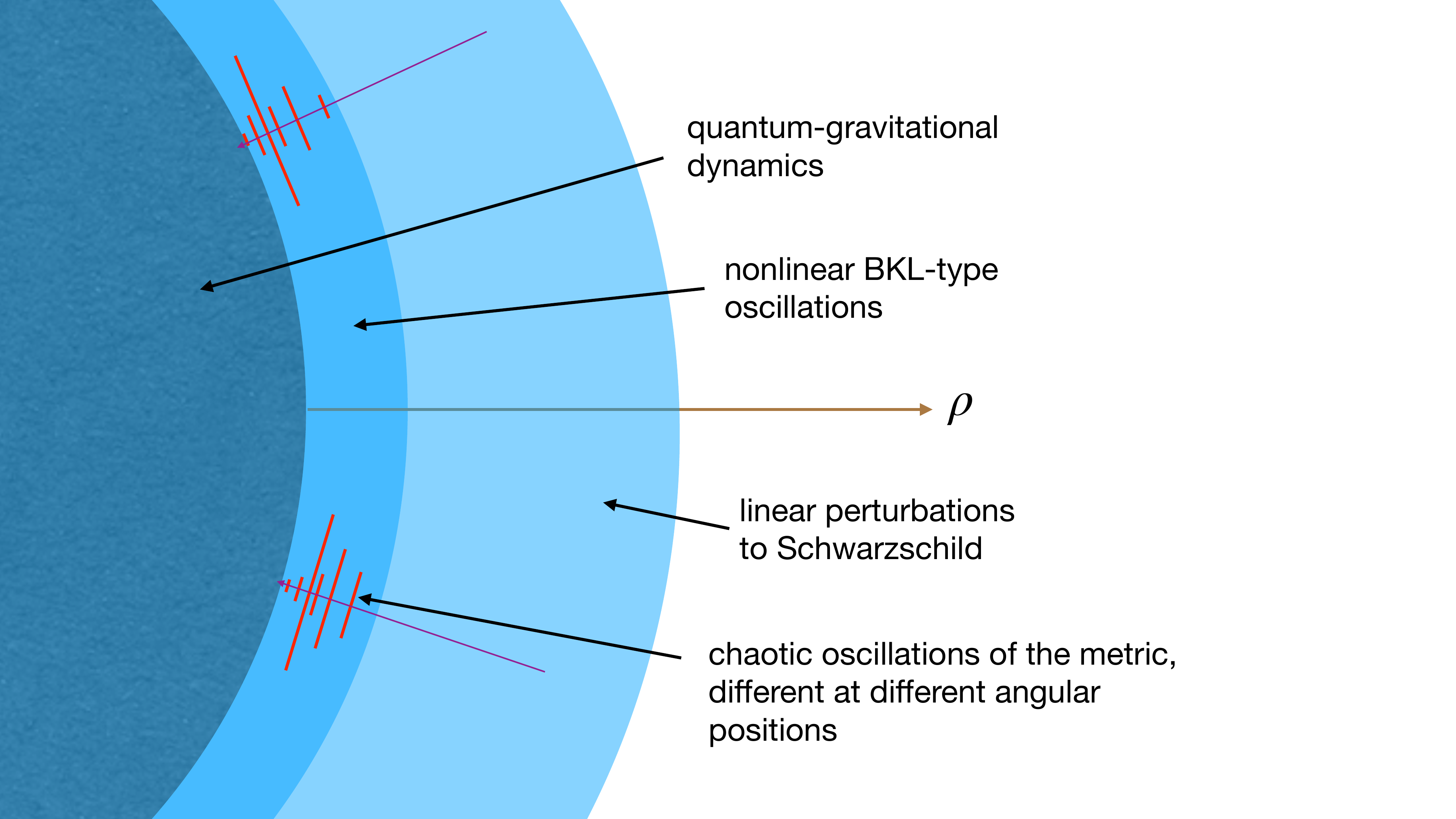}}
  \caption{
     The far region in light blue is described by linearized perturbations to Schwarzschild. In the dark blue band classical Einstein equations still hold, but the perturbations to Schwarzschild are nonlinear, and give BKL-type chaotic structure. The `billiards' type radial evolution ends in runaway squeezing of some compact directions,  leading to the quantum gravitational dynamics in the ECO interior denoted by the grey region. 
    } 
    \label{fig1}
\end{figure}

The above discussion suggests the following description for the structure of an ECO (Figure \ref{fig1}). The interior $r<R_{ECO}$ is stabilized by some quantum gravitational dynamics which we do not explore here. We assume that the exterior $r>R_{ECO}$ is well described by a time-independent solution of the vacuum Einstein  equations $R_{ab}=0$. The interior of the ECO will not be exactly spherically symmetric, so the ECO will have multipole moments of all orders $0<l\lesssim R_{ECO}/l_p$.  For   $r\gg R_{ECO}$  we can linearize the Einstein equations around the Schwarzschild geometry to obtain the metric perturbations $h_{ab}$ describing these multipole moments. Expanding in spherical harmonics, we have
\be
h_{ab}=f_{ab}(r) Y[\theta, \phi]\, .
\label{xone}
\ee
 But as we approach the ECO surface these  perturbations become stronger and  interact,  so we must consider the full nonlinear Einstein  equations. It is this nonlinear region that will be described by BKL-type dynamics, giving the following picture.  Consider a given angular direction $\theta, \phi$, and let $x_1, x_2$ describe the tangent space of these angular directions. As we move inwards along the radial coordinate $\rho$, the metric in the transverse directions $x_1, x_2, t$ expands in one direction while contracting in the other two, with the  rates of change being given by some Kasner indices $p_a$.  At some radius $\rho$  there is a sharp change: we get new indices $p_a$ and new axes of expansion and contraction in the space $x_1, x_2, t$. These sudden changes come at smaller and smaller intervals of   the radial distance $\rho$, until we reach the surface of the ECO and quantum gravitational dynamics takes over.\footnote{There are other contexts where the BKL time $t$ is replaced by a spacelike $\rho$.  BKL-type dynamics has been studied near the  central black hole singularity and  in Bianchi IX type timelike singularities \cite{bkls1}. The CFT duals of Kasner-type   metrics were analyzed in \cite{bkls2}. }

We have recently completed the detailed computations which support  the above picture \cite{fm}. It turns out there is one additional surprise: a phenomenon which occurs  for the case where the evolution parameter is the spacelike coordinate $\rho$ but does not occur for the original BKL story where the evolution parameter $t$ is timelike. 

First consider ECOs in a theory which has only 3+1 spacetime dimensions, and no additional compact directions.  For our radial evolution an analogue of (\ref{three}) would be
\be
ds^2=d\rho^2+  e^{2\beta_1(\rho)}(dx_1+C^1_{2t} x_2 dt)^2+ e^{2\beta_2(\rho)} dx_2^2-e^{2\beta_t(\rho)} dt^2\, .
\label{threeq}
\ee
We define the parameter $u$ through $d\rho= Exp[{\beta_1+\beta_2+\beta_t}]\, du$ and as before, 
 and  denote derivatives $\p_u$ by a prime. We then find
 \be
 \beta''_1= \left (C^1_{2t}\right)^2 e^{4\beta_1}, ~~~ \beta''_2= -\left (C^1_{2t}\right)^2 e^{4\beta_1}, ~~~ \beta''_t= -\left (C^1_{2t}\right)^2 e^{4\beta_1}, 
 \label{fiveq}
 \ee
so we get a reflection of the velocities $\beta'_a$ in an exponential potential wall as in (\ref{six}). But there is an interesting cancellation of signs here. Changing $t$ to $\rho$ actually reverses the sign of the term $\beta''_a$ in Einstein's equations. But the curvature term giving the exponential potential also changes sign when any one of the indices $C^a_{bc}$ is the timelike coordinate $t$. As a result, we find potential walls in (\ref{fiveq}) just like in (\ref{five}), and we get an ergodic billiard dynamics  in the metric as we approach the ECO surface in the radial direction $\rho$. 

But many theories describing ECOs have  extra dimensions; in particular,  string theory  lives in 9+1 spacetime dimensions. Now the coefficients $C^a_{bc}$ capturing curvature can be split into two sets: (i) one of the indices $a,b,c$ is the time $t$ (ii) none of the indices $a,b,c$ is the time $t$. In case (i) we get reflecting walls that bounce the velocities $\beta'_a$ around  ergodically as discussed above. But in case (ii), the sign of the potential changes, so that we get a `cliff' edge instead of a wall. When the velocity $\beta'_a$ reaches a cliff, one gets runaway behavior where the metric in some  directions heads to infinity and in some other directions it heads to zero \cite{fm}. 

This runaway behavior is crucial for relating the above BKL behavior to the quantum gravitational structure of the ECO in string theory. Suppose  a compact circle becomes very small.  Then the low energy limit of the theory ceases to be supergravity, and quantum gravitational effects emerge, as follows:

 (i) A string wrapping this small circle acts as a new very low energy particle in the remaining dimensions.
 
 (ii) A $p$-brane with one direction wrapping the small circle gives a low tension $p-1$ brane in the remaining dimensions. Such a `floppy' extended object has a large degeneracy of states at a relatively low energy.
 
 (iii) With a circle of radius $R$, we have Kaluza-Klein monopole-antimonopole pairs with energy proportional to $R^2$. For 2-charge extremal black holes, we find that the quantum wavefunctional spreads over the space of such pairs, thus capping off the infinite throat of the classical hole and yielding a `fuzzball' \cite{fuzzballs}.

With this, we can now describe the picture we have for an extremely compact ECO.  At large $r$, the geometry is close to the Schwarzschild metric. Closer in, we begin to see the multipole moments of the ECO; these are captured by linear perturbations (\ref{xone}) around the Schwarzschild solution. Our interest lies in what happens as we approach even closer to the ECO surface.  The  physics is still described by the vacuum Einstein equations $R_{ab}=0$, but the linear approximation becomes invalid, and it might appear hard to get any general description of the metric. But the BKL approximation in cosmology gives a natural way to describe this nonlinear region. The variations in the radial direction $\rho$ become much more rapid than those in the tangential directions, so we can focus on a small neighborhood of any point $x$ in the tangential space. As we move towards the ECO surface, the metric takes the Kasner-type form (\ref{oneq}), yielding  rapid expansion in some directions and rapid contraction in others. The axes of expansion and contraction suddenly change at some value of $\rho$, and we get another Kasner-type metric for some radial distance, after which the axes change again. If the total spacetime dimension is larger than $3+1$  (as is the case in most theories with ECOs), then this `billiards' description ends with runaway behavior where a compact direction heads towards zero size. When this size becomes small enough, quantum gravitational effects take over to support the ECO against gravitational collapse.

\b

{\bf A puzzle resolved}

\b
 
The above structure of an ECO resolves a long-standing mystery about alternatives to the semiclassical hole. Far outside the ECO surface, the dynamics will be well described by classical gravity. Inside the ECO the physics must involve novel effects of quantum gravity -- the pressures of normal matter are insufficient to halt gravitational collapse, as seen through results like the Buchdahl theorem.  But it seems odd that the classical dynamics outside the surface  should suddenly change to quantum gravitational dynamics, given that  the curvature of the Schwarzschild solution at the surface is not large.  How exactly does this transition take place? 

The analysis in this essay shows that the transition from the Schwarzschild   region to the quantum gravitational interior is mediated by a naturally arising `in-between' region.  In this region  the classical Einstein equations still hold but  the metric is very chaotic, with large changes from point to point both in the radial and tangential directions. This chaotic dynamics is described by a billiard ball motion of the kind observed by BKL in cosmology. But there is  an interesting difference: some of the reflecting walls of the cosmological case get replaced by cliffs, and reaching a cliff leads to runaway behavior where a compact direction can get squeezed towards zero radius.  This squeezing naturally leads to the structure that has been observed in the construction of black hole microstates in string theory. In these microstates -- called fuzzballs --  different cycles of the compact directions squeeze to zero radius  at different places, so that the spacetime in the fuzzball interior is not a direct product of the compact and noncompact directions. Quantum gravitational effects naturally emerge in this situation, and stabilize the fuzzball against gravitational collapse.  Thus we see that the region with BKL-type dynamics gives an interpolation between the Schwarzschild region far away and the fuzzball interior.

It is fascinating that classical chaotic dynamics automatically cloaks the quantum gravitational region from the smooth spacetime lying farther outside!


\section*{Acknowledgements}

The work of SF   is supported by the University of Waterloo, as well as the  Natural Sciences and Engineering Research Council of Canada, by the Government of Canada through the Department of Innovation, Science and Economic Development and by the Province of Ontario through the Ministry of Colleges and Universities at Perimeter Institute. The work of SDM is supported in part by DOE grant DE-SC0011726.  We  would like to thank Guido Festuccia and Madhur Mehta for  helpful discussions.  


\end{document}